
\input phyzzx
\PHYSREV
\hoffset=0.3in
\voffset=-1pt
\baselineskip = 14pt \lineskiplimit = 1pt
\frontpagetrue
\rightline {Cincinnati preprint July.1994}
\medskip
\titlestyle{\seventeenrm Renormalizability and Quantum Stability
of the Phase Transition in Rigid String Coupled to
Kalb-Ramond Fields I}
\vskip.5in
\medskip
\centerline {\caps M. Awada\footnote*{\rm E-Mail address:
moustafa@physunc.phy.uc.edu}}
\centerline {Physics Department}
\centerline {\it University of Cincinnati,Cincinnati, OH-45221}
\bigskip
\centerline {\bf Abstract}
\bigskip
Recently we have shown that a phase transition occurs
in the leading approximation of the large N limit in rigid strings
coupled to long range  Kalb-Ramond interactions.  The disordered
phase is
essentially the Nambu-Goto-Polyakov string theory while The ordered
phase
is a new theory.  In this part I letter we study the first
sub-leading
quantum corrections of the free rigid string and derive the
renormalization group equation.  We show that the theory is
asymptotically free, thus the extrinsic curvature of the string drops
out
at large distance scales in the disordered phase.  In part II we
generalize
the results of this letter to the interacting theory of rigid strings
with
the long range  Kalb-Ramond interactions.  We derive the renormalized
mass gap equation and obtain the renormalized critical line.  Our
main and
final result is that the phase transition does indeed survive quantum
fluctuations.

\eject

The purpose of this letter is two fold one is to study the free rigid
string [1] to the first sub-leading quantum corrections and show that
the
beta function remains negative, thus implying that the  extrinsic
curvature
becomes irrelevant at large distance scales in the disordered phase.
This is
of utmost importance for the decoupling of ghosts of the quantum
theory [2].
Next the results here will serve as the foundations for solving a
more
complicated problem.  This the problem of showing that the phase
transition
which we proved [3] to occur in the leading approximation of large N
limit
in the theory of rigid strings coupled to the long range Kalb-Ramond
interactions, is stable when we include the sub-leading quantum loop
corrections in Large N.  This would be the topic of part II of this
letter.

 The gauge fixed action of the free rigid string [1] is :
$$I_{gauge-fixed} = \mu_{0}\int d^2\xi\rho + {1\over 2t_{0}}\int
d^2\xi
[\rho^{-1}(\partial^{2}x)^{2} + \lambda^{ab}(\partial_{a}x
\partial_{b}x - \rho\delta_{ab})]\eqno{(1 a)}$$
where t is the curvature coupling constant which is dimensionless.
The partition function is
$$ Z =\int D\lambda D\rho Dx exp(-S_{gauge-fixed})\  .\eqno{(2)}$$
The effective action is obtained by integrating over
$x^{\nu}, \nu=1,...D$
we have:
$$ {S_{0}}_{eff} = {1\over 2t_{0}}[\int d^2\xi (\lambda^{ab}
(-\rho\delta_{ab}) + 2t_{0}\mu_{0}\rho) + t_{0}Dtrln A]\eqno{(3)}$$
where A is the operator
$$ A = \partial^{2}\rho^{-1}\partial^{2} -\partial_{a}
\lambda^{ab}\partial_{b}\  .\eqno{(4)}$$
In the large D limit the stationary point equations resulting
from varying $\lambda$ and $\rho$ respectively are:
$$\rho={t_{0}D\over 2}trG\eqno {(5 a)}$$
$$2t_{0}\mu_{0}-\lambda^{ab}\delta_{ab}=t_{0}Dtr(\rho^{-2}
(-\partial^{2}G))\eqno {(5 b)}$$
where the world sheet Green's function is defined by:
$$ G(\xi,\xi') = <\xi|(-\partial^{2})A^{-1}|\xi'>\eqno{(6)}$$
The stationary points are:
$$ \rho(\xi) = \rho^{*},~~~~~~~~\lambda^{ab}=\lambda^*\delta^{ab}
\eqno{(7)}$$
where $\rho^{*}$ and $\lambda^*$ are constants.  Thus eq.(5a)
becomes the mass gap equation:
$$ 1={Dt_{0}\over 2}\int{d^2p\over (2\pi)^{2}}{1\over p^2+m_{0}^2}
\eqno{(8)}$$
where we define the mass
$$m_{0}^{2}=\rho^*\lambda^*\eqno{(9)}$$
this yields the mass gap equation,
$$ m_{0} = \Lambda e^{-{4\pi\over Dt_{0}}}\eqno{(10 a)}$$
where $\Lambda={1\over a}$ is an U.V. cut-off and $m_{0}$ is now
the bare mass associated with the propagator:
$$ <\partial_{a} x^{\mu}(p)\partial_{a} x^{\nu}(-p)> =
{Dt_{0}\over 2}{\delta^{{\mu}{\nu}}\over p^2+m_{0}^2}
\  .\eqno{(10 b)}$$
On the other hand eq(5b) yields the string tension
renormalization condition:
$$\mu_{0}= {D\over 8\pi}{\Lambda^{2}\over \rho^*}
\eqno{(11)}$$
eq. (10a) agrees exactly with the one loop result.
\bigskip
{\bf II-The Loop Corrected Gap Equation and the RG Equation}

In mean field theory i.e leading order in ${1\over D}$,
the relevant propagator is equation (10 b).  In the sub-leading
correction to mean field theory, the quantum fluctuation
imply a new term corresponding to the self-energy of the
$\partial_{a} x^{\mu}$-field
$$ <\partial_{a} x^{\mu}(p)\partial_{a} x^{\nu}(-p)> =
{Dt_{0}\over 2}{\delta^{{\mu}{\nu}}\over (p^2+m_{0}^2
+ {1\over D}\Sigma(p))}\  .\eqno{(12)}$$
The new contribution $\Sigma (p)$ arises from fluctuations of the
Lagrange multipliers $\lambda_{ab}$ and $\rho$ where the fluctuations
$\sigma_{ab}$ and $\eta$ are defined by:
$$ \lambda_{ab} = \lambda^{*}\delta_{ab} +
i{1\over \sqrt{(D/2)}}\sigma_{ab}$$
$$\rho = \rho^{*}(1 + i{\rho^{*}\over \sqrt{(D/2)}}\eta)\
.\eqno{(13)}$$
Expanding the effective action (3) in powers of $\sigma_{ab}$ and
$\eta$,
it is straightforward to extract the $\sigma_{ab}$ and
$\eta$ propagators (Fig (1)):
$$ \Pi_{ab|cd}(p) = {1\over 2}(\delta_{ac}\delta_{bd} +
\delta_{ad}\delta_{bc})\pi(p^2)\eqno{(14 a)}$$
$$\pi(p^2) =
\int {d^{2}k\over (2\pi)^{2}} {1\over (k^2 + m_{0}^{2})
((p + k)^2 + m_{0}^{2})}\eqno{(14 b)}$$
$${\tilde \pi}(p^2) =
2\int {d^{2}k\over (2\pi)^{2}}{(k.(k+p))^2\over (k^2 + m_{0}^{2})
((p + k)^2 + m_{0}^{2})}\  .\eqno{(14 c)}$$
It is obvious from (14c) that the $\eta$ propagator has quadratic and
logarthimic divergences and therefore needs regularization before
attempting to compute the self energy.  Eqs.(14) can be exactly
computed and one finds :
$$\pi(p^2) = {1\over 2\pi p^2\xi}ln{\xi+1\over \xi-1}\eqno{(15 a)}$$
$${\tilde \pi}(p^2) = \pi^{*}(p^2) + {\Lambda^2\over 2\pi}
- {1\over 4\pi}(p^2 + 4m_{0}^2)ln{\Lambda^2\over m_{0}^2}\eqno{(15
b)}$$
where the finite regularized propagator is :
$$\pi^{*}(p^2) = 2m_{0}^4\pi(p^2)+ {p^2\over 4\pi}
(\xi ln{\xi+1\over \xi-1} - 1)\eqno{(15 c)}$$
where
$$\xi = (1+{4m_{0}^2\over p^2})^{{1\over 2}}\  .\eqno{(15 d)}$$

The self energy $\Sigma$ can be computed from the diagrams of
Fig(2).  These diagrams are of order ${1\over D}$ and represent
the quantum fluctuations:
$$\Sigma(p) = \int{d^2k\over (2\pi)^2}
{\pi^{-1}(k^2)\over ((p + k)^2 + m_{0}^2)}+{1\over 2}\int{d^2k\over
(2\pi)^2}
{{\pi^*}^{-1}(k^2)(k.(k+p))^2\over ((p + k)^2 + m_{0}^2)}$$
$$ - \int {d^2k\over (2\pi)^2} \int{d^2q\over (2\pi)^2} {\pi^{-1}(0)
\over (q^2 + m_{0}^2)^{2}}
{\pi^{-1}(k^2)\over ((q + k)^2 + m_{0}^2)}$$
$$ - {1\over 2}\int {d^2k\over (2\pi)^2} \int{d^2q\over (2\pi)^2}
{\pi^{-1}(0)\over (q^2 + m_{0}^2)^{2}}
{{\pi^*}^{-1}(k^2)(k.(k+q))^2\over ((q + k)^2 + m_{0}^2)}
\  .\eqno{(16)}$$
A Taylor expansion of the self energy about zero momentum leads
to mass and wave function renormalizations and a remaining piece
${\tilde \Sigma}$ which must be finite for the theory to be
renormalizable.  The propagator now reads:
$${Z\over (p^2 + m^2 + {1\over D}{\tilde \Sigma_{finite}(p)})}
\  .\eqno{(17)}$$
where
$$ Z = 1 - {1\over D} \Sigma'(0)\eqno{(18 a )}$$
is the wave function renormalization and
$$m^2 =m_{0}^2 + {1\over D}(\Sigma(0) -m_{0}^2
\Sigma'(0))\eqno{(18 b )}$$
is mass renormalization.
In order to calculate the finite regularized self energy we need to
simplify
further Eq.(16).  Using (14) we can replace zero-momentum insertions
of the
$\sigma$ and $\rho$ fields and rewrite (16) as:
$$\Sigma(p) = \Sigma_{1}(p) + \int{d^2k\over (2\pi)^2}
{\pi^{-1}(k^2)\over ((p + k)^2 + m_{0}^2)} + {1\over 2}\pi(0)^{-1}
\int{d^2k\over (2\pi)^2}\pi^{-1}(k^2){\partial\over
m_{0}^2}\pi(k^2)$$
$$ +{1\over 2}\int{d^2k\over (2\pi)^2}
{{\pi^*}^{-1}(k^2)(k.(k+p))^2\over ((p + k)^2 + m_{0}^2)} +
{1\over 4}\pi(0)^{-1}\int{d^2k\over (2\pi)^2}{\pi^*}^{-1}(k^2)
{\partial\over m_{0}^2}{\pi^*}(k^2)$$
$$ + {1\over 2}m_{0}^2 log{\Lambda^2\over m_{0}^2}\int{d^2k\over
(2\pi)^2}
{\pi^*}^{-1}(k^2) -{1\over 2}\int{d^2k\over (2\pi)^2} k^2
{\pi^*}^{-1}(k^2)
\eqno{(19)}$$
where $\Sigma_{1}(p)$ is a finite piece of the self energy.  From the
explicit form of the propagators in (15) one can derive the
following identities:
$${\partial\over m_{0}^2}\pi(k^2) = {-2\over k^2 \xi^{2}}(\pi(k^2)
+\pi(0))
\eqno{(20 a)}$$
$${\partial\over m_{0}^2}{\pi^*}(k^2) = ({1\over m_{0}^2}-
{2\over k^2 \xi^{2}}){\pi^*}(k^2) -{2\over k^2 \xi^{2}}{\pi^*}(0)
+2 m_{0}^2 \pi(k^2) - {k^2\over 4\pi m_{0}^2 \xi}ln{\xi+1\over \xi-1}
-{1\over 2\pi \xi^{2}}
\  .\eqno{(20 b)}$$
Inserting (20) into (19) we obtain:
$$\Sigma(p) = \Sigma_{1}(p) + {\Lambda^2\over 4} - {3\over 2}
m_{0}^2 log{\Lambda^2\over 4m_{0}^2}
+\int{d^2k\over (2\pi)^2}{\pi^{-1}(k^2)\over ((p + k)^2 + m_{0}^2)} -
\int{d^2k\over (2\pi)^2}{\pi^{-1}(k^2)\over (k^2 + 4m_{0}^2)}$$
$$+{1\over 2}\int{d^2k\over (2\pi)^2}
{{\pi^*}^{-1}(k^2)(k.(k+p))^2\over ((p + k)^2 + m_{0}^2)}
-{1\over 2}\int{d^2k\over (2\pi)^2} k^2 {\pi^*}^{-1}(k^2) -
{m_{0}^2\over 2}\int{d^2k\over (2\pi)^2}{\pi^*}^{-1}(k^2)$$
$$ +{1\over 2} m_{0}^2 log{\Lambda^2\over m_{0}^2}\int{d^2k\over
(2\pi)^2}
{\pi^*}^{-1}(k^2) + m_{0}^4\int{d^2k\over (2\pi)^2}{{\pi^*}^
{-1}(k^2)\over (k^2 + 4m_{0}^2)}$$
$$ +2\pi m_{0}^4\int{d^2k\over (2\pi)^2}{\pi^*}^{-1}(k^2)\pi(k^2)
 - {1\over 4}\int{d^2k\over (2\pi)^2}{\pi^*}^{-1}(k^2){k^2\over \xi}
ln{\xi+1\over \xi-1}\  .\eqno{(21)}$$

Now we are in a good position to regularize (21).  We proceed
by applying the SM regularization scheme [4] where one subtracts the
highest powers of the integration variable appearing in the Taylor
expansion of the integrands in (21).  The divergences in (21) occur
at
the zeroth and first order expansion.  Using the asymptotic limits of
the propagators (15):
$$\pi_{asy}(p^2) = {1\over 2\pi p^2} log{p^2\over m_{0}^2}\eqno{(22
a)}$$
$$\pi^{*}_{asy}(p^2) = {p^2\over 4\pi} log{p^2\over m_{0}^2}\eqno{(22
b)}$$
we obtain the following finite regularized self energy:
$$\Sigma_{finite}(p) = \Sigma(p) + m_{0}^2 log{\Lambda^2\over
m_{0}^2}
(1-{1\over 2} I_{0}) -{m_{0}^2\over 4} I_{0} - {1\over 4}(p^2 +
m_{0}^2) I_{0}\eqno{(23 a)}$$
where
$$I_{0}({\Lambda^2\over m_{0}^2}) = \int{d^2k\over
(2\pi)^2}{4\pi\over
k^2 log{k^2\over m_{0}^2}}
= loglog{\Lambda^2\over m_{0}^2}\  .\eqno{(23 b)}$$
We have arranged $\Sigma_{finite}(p)$ in the above form so as it is
easy to see how the above result (23) can be generalized to the
interacting
theory of rigid string with the long range Kalb-Ramond interactions
presented
in part II.  We can now read from (23) the mass and wave function
renormalizations:
$$ m^2 = m_{0}^2[1 - {1\over D}(log{\Lambda^2\over m_{0}^2}
(1- {1\over 2}I_{0}) -{1\over 4} I_{0})]\eqno{(24 a)}$$
$$ Z = 1 - {1\over 4D}I_{0}\  .\eqno{(24 b)}$$

To obtain the beta function we hold $m^2({\Lambda^2\over m_{0}^2},t)$
fixed we obtain:
$$\beta(t) = -({\partial log m^2\over \partial t})^{-1}$$
$$\beta(t^*)= -{t^*}^2(1 - {1\over 2D})(1 + {1\over 2D}
({1\over 2}t^* - logt^*))\eqno{(25)}$$
where we have used the mass gap equation (10a) and defined
$t^*={Dt\over 8\pi}$.  Eq.(25) is the main result
of this letter. It shows that the free rigid string theory is
asymptotically free to the sub-leading quantum corrections in  the
large D limit.  The critical point is still the ultra violet fixed
point
$t=t_{c}=0$.  Therefore at large distance scales in the disordered
phase
$t>t_{c}$ the extrinsic curvature of the string becomes irrelevant
and the
resulting classical limit is the usual Nambu-Goto string which is
free of
ghosts.  The rigid string is a higher derivative theory and therefore
can
have the typical pathologies, namely classical runaway solutions and
ghosts
appearing in the perturbative propagator.  However the
non-perturbative
(in the coupling t) result (25) shows that the regulated quantum
theory
is consistent.  the  extrinsic curvature term only survives at very
short
distance scales and does not affect the poles of the propagator,
which is
a large distance property.  Indeed in [2], and [5] we have proved
that
rigid QED which is a higher derivative theory is a consistent quantum
theory
both in the leading and sub-leading orders in large N limit and that
the
phase transition survives sub-leading quantum fluctuations.
\eject

{\bf Acknowledgement}

 I am very grateful to Prof. A. Polyakov for his constant
encouragement and
extensive support over the last year and a half and for suggesting
that we
address the quantum stability of the phase transition both in the
model of
rigid QED and that of rigid strings coupled to long range
interactions [6].
I am also grateful to Prof. Y. Nambu for his extensive support,
encouragement
and long discussions over the last two years without which we could
not have
gone far in our investigations. I also thank P.Ramond, C. Thorn, and
Z. Qui
for constructive discussions and suggestions. Finally my gratitude to
my
friend D. Zoller for a fruitful and devoted collaboration over the
years.

{\bf References}

\item {[1]} A. Polyakov, Nucl. Phys. B268 (1986) 406
; A. Polyakov,  Gauge fields, and Strings,
Vol.3, harwood academic publishers
\item {[2]} M. Awada and D. Zoller, Phys.Lett B325 (1994) 119
\item {[3]} M. Awada and D. Zoller, Phys.Lett B325 (1994) 115
\item {[4]} A. Polyakov,  Gauge fields, and Strings,
Vol.3, harwood academic publishers, J.Orloff and R.Brout, Nucl.
Phys. B270 [FS16],273 (1986), M. Campostrini and P.Rossi, Phys.
Rev.D 45, 618 (1992) ; 46, 2741 (1992), H. Flyvberg, Nucl. Phys.
B 348, 714, (1991).
\item {[5]} M. Awada, D. Zoller, and J. Clark, Cincinnati preprint
June -1 -(1994)
\end